# Coding Side-Information for Implementing Cubic Transformation

B. Prashanth Reddy

**Abstract**
This paper considers the implementation of the cubic public-key transformation, a public-key cryptographic scheme that requires sending of additional side-information. A coding scheme for the side-information, based on the residue number system, is presented. In the conventional one-to-one encryption mapping also, such coding can be used to send additional like that of watermarking, which could be used to detect man-in-the-middle attacks or used for authentication.

## 1 Introduction

Cybersecurity must deal not only with communications security [1]-[8] but also access controls, random sequences [9]-[17], secret-sharing [18]-[25], P2P networks [26]-[28], and cryptocurrency [29]-[31]. NSF's National Coordination Office for Networking and Information Technology Research and Development has identified "moving target by increasing the cost of attack, such as by making the security environment dynamic and therefore harder to predict or less susceptible to prolonged attack" as one of the most important methods of dealing with potential attackers. It is in that spirit we are considering a hybrid system of cryptography that has both public-key encryption as well as side-information that needs to be sent. Specifically, we consider the cubic transformation [9], which is one-to-one only if additional side-information is sent along with the encrypted data.

In general, a limited many-to-one mapping can be represented by $f(m_i) = c$, $i \geq 2$, where $m_i$ are the various messages that lead to the same encrypted sequence $c$ (Figure 1). When $i=2$, we have the basis of oblivious transfer protocol [18],[19].

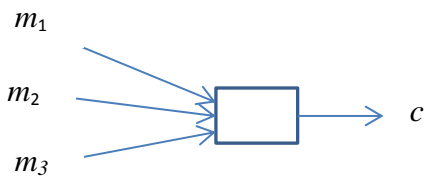

Figure 1. Many-to-one mapping for cryptography using side information of $i$

To convert $f(m_i) = c, i \geq 2,$ into a one-to-one mapping, one needs to use it in a slightly different form where in addition to $c$, the correct value of $i$ is also transmitted to the recipient as side-information, quite like the piggy bank cryptographic paradigm [19].

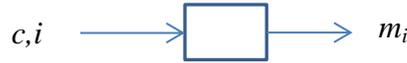

Figure 2. Inverse mapping

We consider several examples of the transformation $f(m_i) = c, i=3$, that sheds light on the issues involved in an implementation. Specifically, we propose a residue number representation of the indexing information. Enhanced indexing can also be used to send additional information like that of watermarking, which could be used to detect man-in-the-middle attacks in standard one-to-one mapping schemes.

**2 Coding indexing information for transformation mod prime**

We propose the use of the residue number system (RNS) to code the indexing information which is labeled 0, 1, 2, .. . This will increase the size of the words minimally. Let us assume that the number c is mod s and the index is mod r. Then $c,i$ will be mapped to the number $T$ and

$T \bmod s = c$
$T \bmod r = i$

As example the cubic transformation mod 13 leads to the mapping as in the table below (s=13 and r =3):

Table 1: RNS coding scheme with M=13x3=39

| m | $c=m^3$ | c,i indexed | c,i in RNS |
|---|---|---|---|
| 1,3,9 | 1 | 1,0; 1,1;1,2 | 27,1,14 |
| 2,5,6 | 8 | 8,0;8,1;8,2 | 21,34,8 |
| 4,10,12 | 12 | 12,0;12,1;12,2 | 12,25,38 |
| 7,8,11 | 5 | 5,0;5,1;5,2 | 18,31,5 |

As explanation, m=9, leads to the cube 1, which is the third, largest, indexed, therefore, with 0, corresponding to 1,2. In RNS mod 39, this is the unique number 14, which is what the transmitted code will be.

At the receiver, 14 will be first transformed to 1,2, for which the three cube roots of 1 will be computed, leading, in turn, to 9 as the message. The general framework is shown in Figure 3.

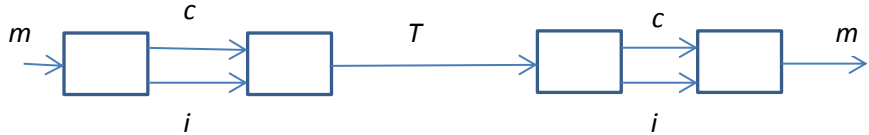

Figure 3. Schematic for the signal flow

Consider message m in $z_n$, with encryption function $c = m^3$ mod p, where p=3k+1 is a prime and p=3 mod 4. This transformation is not one-to-one for p-1 and 3 have at least one common divisor. It is further assumed that 3 divides the Euler's totient $\varphi(n)$, but 9 does not. Thus three different values of m will map to same c. The cube roots are 1, $\alpha$ and $\alpha^2$ because if $\alpha$ is a root, so is its square. The three cube roots of 1 may be obtained by solving the equation

$$\alpha^3 - 1 = 0 \tag{1}$$

Excluding 1 of (1), the other two are obtained by solving $\alpha^2 + \alpha + 1 = 0$, which is possible if the square root is of $\sqrt{p-3}$ exists. The roots can also be written as $\alpha$ and $\alpha^2$, so the value of $\alpha$ may be equivalently expressed as

$$\alpha = \frac{-1 + (p-3)^{\frac{p+1}{4}}}{-1 - (p-3)^{\frac{p+1}{4}}} \tag{2}$$

Ensuring p-1 not be divisible by 9, one can get all the three $c^{\frac{1}{3}}$ values by exponentiation

$$c^{\frac{1}{3}} = \begin{cases} c^{\frac{p+2}{9}}, & \text{if } \varphi(p) \text{ or } (p-1) \bmod 9 = 6 \\ c^{\frac{2p+1}{9}}, & \text{if } \varphi(p) \text{ or } (p-1) \bmod 9 = 3 \end{cases} \tag{3}$$

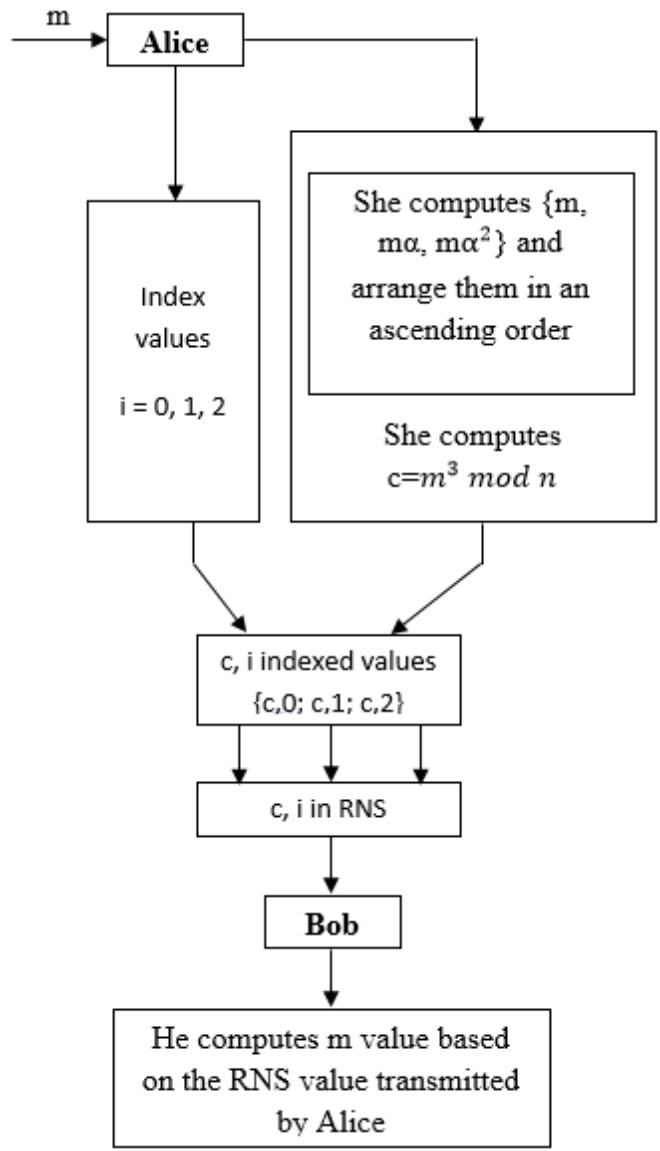

Figure 4. Communication between Alice and Bob

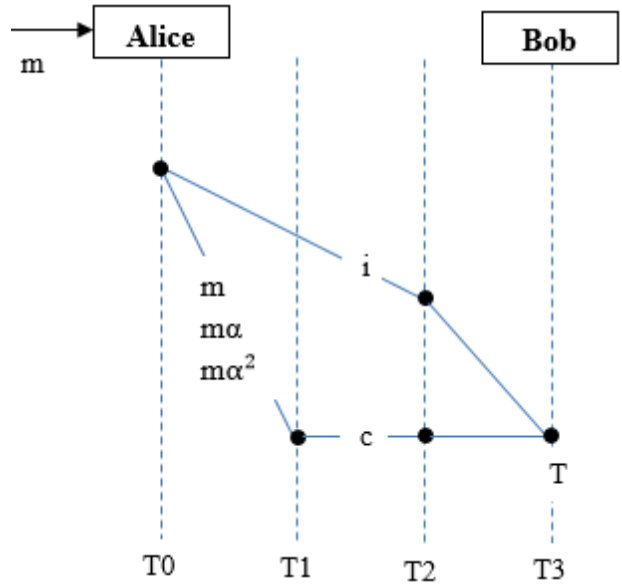

Figure 5. Timing diagram represents communication between Alice and Bob

Table 2: Communication between Alice and Bob for CT for the case of modulo 907

| Alice | Bob |
|---|---|
| Let message m chosen by Alice is 8<br><br>After computing $\{m, m\alpha, m\alpha^2\}$ and arranging them in an ascending order, she assigns index values accordingly for the obtained cubic roots $\{8, 243, 692\}$<br><br>Alice computes $c=8^3 \bmod 907 = 512$ along with for the above computed roots and represent these message values in RNS based on the index values<br><br>Alice finally transmits desired RNS value to Bob | Bob computes m value based on the RNS value transmitted by Alice |

## 3 When the modulus is composite

Alice chooses large composite number n and publishes it. While decrypting a cipher text Bob requires c and rank information along with the publicly published composite number n. Now we generalize the previous analysis to apply composite moduli. As in RSA, we perform cubic transformation with a composite number n, where n = product of two primes p and q. φ(n) (Euler's totient) is product of (p −1) and (q −1) is divisible by 3 but not 9.

$$c^{\frac{1}{3}} = \begin{cases} c^{\frac{(\varphi(n))+3}{9}}, & \text{if } \varphi(n) \bmod 9 = 6 \\ c^{\frac{2(\varphi(n))+3}{9}}, & \text{if } \varphi(n) \bmod 9 = 3 \end{cases} \quad (4)$$

Both Alice and Bob know the factors of n, but the eavesdropper does not. Also, α is published for public and factors of φ(n) are private. Ensuring φ(n) not be divisible by 9, one can get all the three $c^{\frac{1}{3}}$ values by exponentiation.

If we consider the example of n= 533, one must first find two of its prime factors p=13 and q=41. This is done directly by the Chinese Remainder Theorem (CRT), by solving $\alpha^3-1=0$ separately for the two moduli 13 and 41.

To find α only one new solution is required. Using the square modulo operation, Alice finds the cube roots for $\alpha_p$ = {2, 4} and $\alpha_q$ = {1} and combining together and applying CRT,

$$\alpha = \left(\alpha_p \cdot q \cdot \|q\|_p^{-1} + \alpha_q \cdot p \cdot \|p\|_q^{-1}\right) \bmod pq \quad (5)$$

where $\|p\|_q$=p mod q. Since $13^{-1}$ mod 41= 19 and $41^{-1}$ mod 13= 7, $\alpha_1$=42 and $\alpha_2$=329. It is assumed that Alice and Bob have chosen in advance to use α = 42. It's now time for Alice to choose a message (m = 8) to Bob. As in the previous case, she finds its companions by multiplying it successively by α = 42, thus obtaining {8,254,336}.

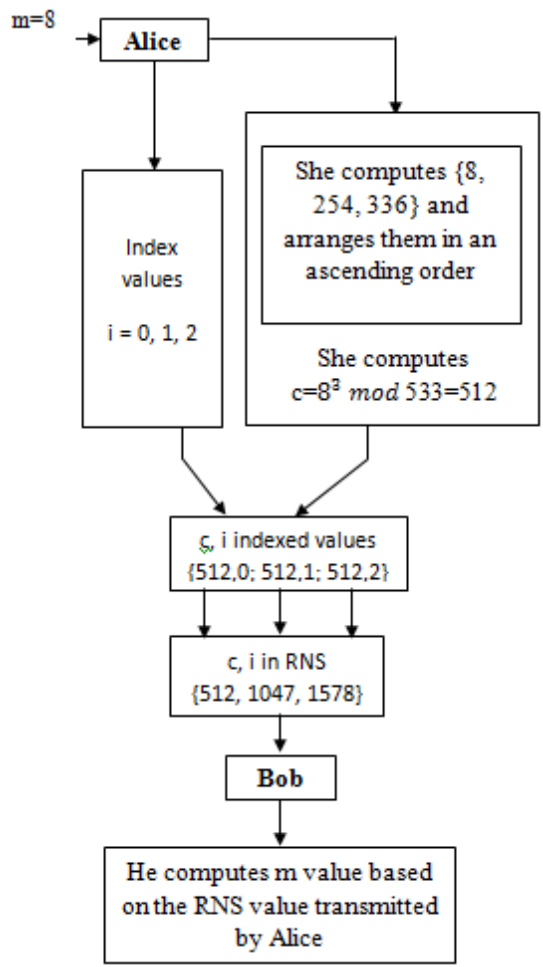

Figure 6. Communication between Alice and Bob

Table 3: Mapping for message m

| m | c = m$^3$ | c, i indexed | c, i in RNS |
|---|---|---|---|
| 1, 42, 165 | 1 | 1,0; 1,1; 1,2 | 1, 534, 1067 |
| 2, 84, 330 | 8 | 8,0; 8,1; 8,2 | 8, 541, 1074 |
| 4, 127, 168 | 64 | 64,0; 64,1; 64,2 | 64, 597, 1130 |
| 8, 254, 336 | 512 | 512,0; 512,1; 512,2 | 512, 1047, 1578 |

Table 4: Communication between Alice and Bob for CT modulo composite

| Alice | Bob |
|---|---|
| Let message m chosen by Alice is 8<br><br>After computing $\{m, m\alpha, m\alpha^2\}$ and arranging them in an ascending order, she assigns index values accordingly for the computed cubic roots {8, 254, 336}<br><br>Alice computes $c=8^3 \bmod 533 =512$ along with the computed roots and represent these message values in RNS based on the index values<br><br>Alice finally transmits desired RNS value to Bob | Bob computes m value based on the RNS value transmitted by Alice |

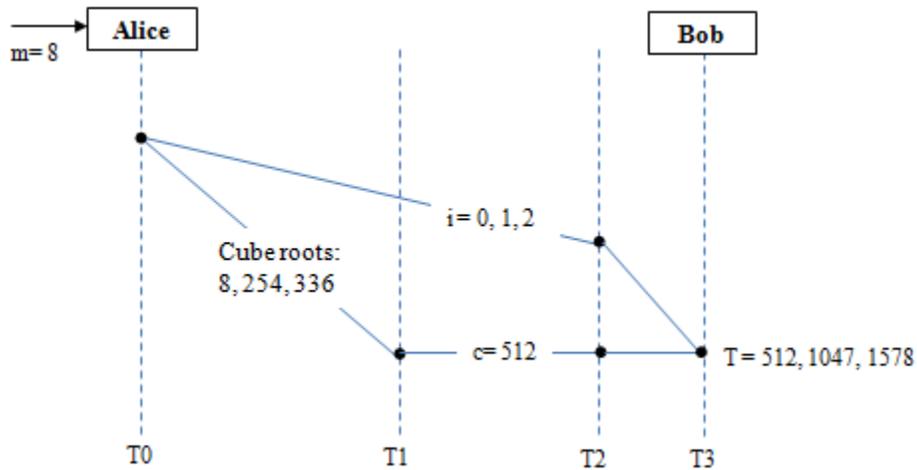

Figure 7. Timing diagram represents communication between Alice and Bob

*3.1 Composite number divisible by 9*

When φ(n) is divisible by 9, each number will now have nine cube roots that are the solutions of the equation given below:

$$\alpha^9 - 1 = 0 \quad (6)$$

or, equivalently, by solving:

$$(\alpha - 1)(\alpha^2 + \alpha + 1)(\alpha^6 + \alpha^3 + 1) = 0 \quad (7)$$

The cube roots of 1 can be indexed 1 through 9 by their numerical value. Any message needs to be multiplied by the 9 cube roots of 1 to find its location in the indexed set of 9 values.

One of the important criteria to complete cubic transformation is to pick a composite number n such that $\varphi(n)$ is divisible by 9.

Consider, for example, that n=679. Alice decomposes finds two of its prime factors p=7 and q=97. By using CRT Alice finally computes 3 roots for $\alpha_p$ = {1, 2, 3} and another 3 roots $\alpha_q$ = {1, 2, 4}.

$\alpha_1$ = (1.97.14+1.7.6) mod 679 = 1
$\alpha_2$ = (2.97.14+1.7.6) mod 679 = 538
$\alpha_3$ = (3.97.14+1.7.6) mod 679 = 389
$\alpha_4$ = (1.97.14+2.7.6) mod 679 = 99
$\alpha_5$ = (2.97.14+2.7.6) mod 679 = 2
$\alpha_6$ = (3.97.14+2.7.6) mod 679 = 487
$\alpha_7$ = (1.97.14+4.7.6) mod 679 = 197
$\alpha_8$ = (2.97.14+4.7.6) mod 679 = 100
$\alpha_9$ = (3.97.14+4.7.6) mod 679 = 585

Hence, the 9 roots obtained will be {1, 538, 389, 99, 2, 487, 197, 100, 85}. Let message m that Alice wants to send to Bob is 233. Multiplying 233 with above obtained 9 roots and arranging them in ascending order, Alice obtains {39, 78, 214, 233, 330, 408, 466, 505, 600}.

Alice sends c = $233^3$ mod 679=246 to Bob. Bob computes m value based on the RNS value transmitted by Alice.

Table 5: Mapping for message m performing CT divisibility by 9 modulo $\varphi(n)$

| m | c = $m^3$ | c, i indexed | c, i indexed |
|---|---|---|---|
| 39, 233, 330 | 246 | 246,0; 246,1; 246,2 | 246, 925, 1604 |
| 214, 408, 505 | 337 | 337,0; 337,1; 337,2 | 1695, 337, 1016 |
| 78, 466, 660 | 610 | 610,0; 610,1; 610,2 | 1968, 610, 1289 |

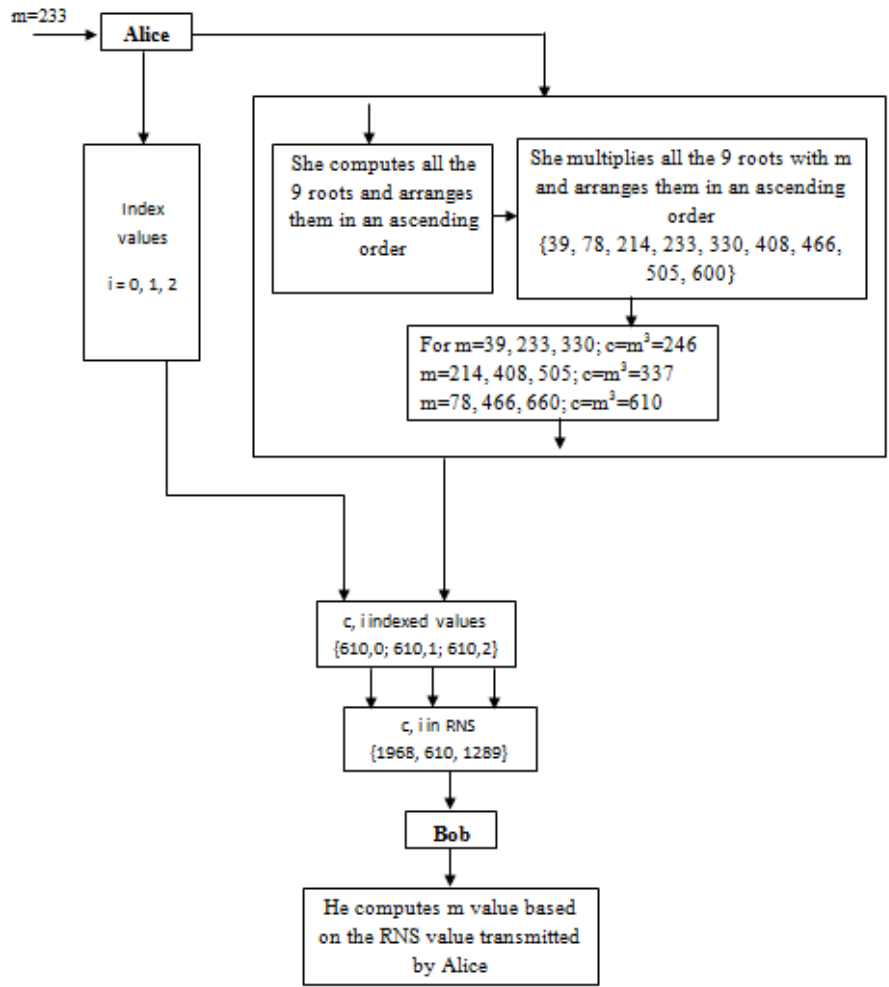

Figure 8. Timing diagram represents communication between Alice and Bob

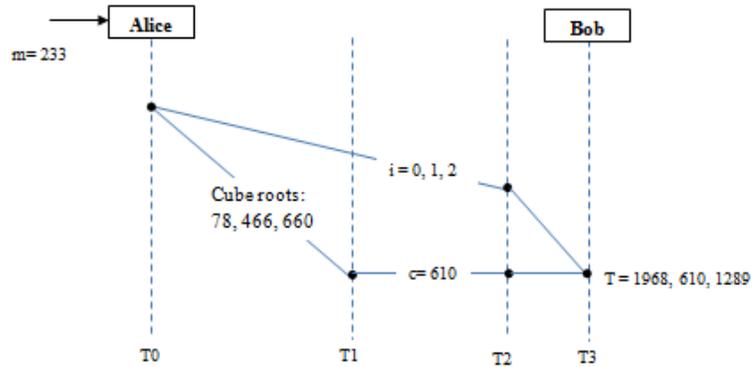

Figure 9. Timing diagram represents communication between Alice and Bob

## 4 Conclusions

This paper discusses the implementation of the cubic public-key transformation to convert it into a one-to-one mapping suitable for message encryption. A coding scheme for the side-information, based on the residue number system, is presented and illustrated with examples.

The idea of the coded side-information can also be applied to conventional one-to-one encryption mapping. Such side-information can then serve as a cover for the encrypted message. In such a case, coding can be used to send additional like that of watermarking, which could be used to detect man-in-the-middle attacks or for authentication.

**REFERENCES**
[1]  O. Renn, Risk governance: coping with uncertainty in a complex world. Earthscan, 2008.
[2]  F. Rocha, S. Abreau, M. Correia, The final frontier: confidentiality and privacy in the cloud. IEEE Computer, vol. 44, September 2011.
[3]  R. Lukose K. Nagaraja J. Pruyne B. Richard S. Rollins D.S. Milojicic,V. Kalogeraki and Z. Xu. Peer-to-peer computing. Technical Report HPL-2002-57R1, HP Laboratories, Palo Alto, CA, USA, 2003.
[4]  B. Wellman, The rise (and possible fall) of networked individualism. Connections, 2002.
[5]  S. Kak, The Nature of Physical Reality. Peter Lang, New York, 1986, 2011.
[6]  B. Wellman, Computer networks as social networks. Science 293: 2031-2034, 2001.
[7]  S. Kak, The Architecture of Knowledge. CSC, New Delhi, 2004.
[8]  J. Garay, R. Gennaro, C. Jutla, T. Rabin, T. Secure distributed storage and retrieval. Theoretical Computer Science, volume 243, issue 1-2, pages 363 – 389, 2000.

[9] S. Kak, A cubic public-key transformation. Circuits, Systems and Signal Processing 26: 353-359, 2007.

[10] A. Kolmogorov. Three approaches to the quantitative definition of information, Problems of Information Transmission, 1, 1-17, 1965.

[11] S. Kak. Encryption and error-correction using d-sequences. IEEE Trans. On Computers, vol. C-34: 803-809, 1985.

[12] D. Eastlake 3rd, S. Crocker, J. Schiller, Randomness Recommendations for Security. Network Working Group, MIT, 1994.

[13] S. Kak and A. Chatterjee. On decimal sequences. IEEE Transactions on Information Theory, IT-27: 647 – 652, 1981.

[14] S. Kak, The initialization problem in quantum computing. Foundations of Physics 29: 267-279, 1999

[15] R. Landauer, The physical nature of information. Physics Letters A 217: 188-193, 1996

[16] S. Kak, A three-stage quantum cryptography protocol. Foundations of Physics Letters 19: 293-296, 2006.

[17] S. Kak, Quantum information and entropy. Int. Journal of Theo. Phys. 46: 860-876, 2007.

[18] A. J. Paul. C and A. Scott. Handbook of Applied Cryptography, Library of Congress Cataloging-in-Publication Data, 1965.

[19] S. Kak, The piggy bank cryptographic trope. Infocommunications Journal 6: 22-25, 2014.

[20] A. Shamir, How to Share a Secret, Communications of the ACM, 22: 612-613, 1979

[21] G.R. Blakely, Safeguarding Cryptographic Keys, Proceedings of the National Computer Conference, American Federation of Information Processing Societies Proceedings, 48: 313-317, 1979

[22] A. Parakh, S. Kak, Online data storage using implicit security. Information Sciences, 179: 335-341, 2009.

[23] A. Parakh, S. Kak, Space efficient secret sharing for implicit data security. Information Sciences, 181: 335-341, 2011.

[24] M. Rabin, Digitalized signatures and public key functions as intractable as factoring. Tech. Rep. MIT/LCS/TR-212, MIT (1979)

[25] S. Even, O. Goldreich, A. Lempel, A randomized protocol for signing contracts. Comm. of the ACM 28: 637-647 (1985)

[26] L. Washbourne, A survey of P2P Network security. arXiv:1504.01358, 2015.

[27] R. Gunturu, Survey of Sybil attacks in social networks. arXiv:1504.05522, 2015.

[28] S. Gangan, A review of man-in-the-middle attacks. arXiv:1504.02115, 2015.

[29] I. Miers and C. Garman., M. Green., A. D. Rubin, Zerocoin: Anonymous distributed e-cash from bitcoin, IEEE Symposium on Security and Privacy, 2013.

[30] P. C. P. Bhatt, What's new in computers: Cryptocurrencies: an introduction, Resonance, 19:549-569, 2014.

[31] J. Bohr. and M. Bashir. Twelfth Annual Conference on Privacy, Security and Trust (PST), IEEE, Who Uses Bitcoin? An exploration of the Bitcoin community, 2014.